\documentclass[12pt,preprint]{aastex}

\newcommand\5{{\footnotesize V}}
\newcommand\4{{\footnotesize IV}}

\newcommand\1{{\footnotesize I}}
\newcommand\lam{$\lambda$}
\newcommand{\sref}[1]{\S\ref{#1}}
\newcommand{\eref}[1]{(\ref{#1})}
\newcommand{\ea}{{et al.}}

\newcommand\kms{\ensuremath{\mbox{km s}^{-1}}}
\newcommand\vsini{\ensuremath{v\sin{i}}}
\newcommand{\p}{\ensuremath{\phantom{:}}}

\newcommand{\ph}[1]{\ensuremath{\phantom{#1}}}

\slugcomment{Accepted 2004 February 3}
\shorttitle{Terminal velocities of luminous, early-type SMC stars}
\shortauthors{Evans et al.}

\received{2004 January 10}
\accepted{2004 February 3}
\begin{document}


\title{Terminal velocities of luminous, early-type SMC stars}


\author{ C. J. Evans\altaffilmark{1},
         D. J. Lennon\altaffilmark{1},
         C. Trundle\altaffilmark{1,2},
         S. R. Heap\altaffilmark{3}, 
         D. J. Lindler\altaffilmark{3}}

\altaffiltext{1}{Isaac Newton Group of Telescopes, 
                 Apartado de Correos 321, 
                 38700 Santa Cruz de la Palma, 
                 Canary Islands, 
                 Spain.}
\altaffiltext{2}{Dept. of Pure and Applied Physics, 
                 Queen's University of Belfast, 
                 Belfast, 
                 BT7 1NN,  
                 N. Ireland, 
                 UK.}
\altaffiltext{3}{Laboratory for Astronomy and Solar Physics, 
                 Code 681, 
                 NASA Goddard Space Flight Center, 
                 Greenbelt, 
                 MD 20771.}

\begin{abstract}
Ultraviolet spectra from the Space Telescope Imaging Spectrograph
(STIS) are used to determine terminal velocities for 11 O and B-type
giants and supergiants in the Small Magellanic Cloud (SMC) from the
Si~\4 and C~\4 resonance lines.  Using archival data from observations
with the Goddard High-Resolution Spectrograph and the {\it
International\,Ultraviolet\,Explorer} telescope, terminal velocities
are obtained for a further five B-type supergiants.  We discuss the
metallicity dependence of stellar terminal velocities for supergiants,
finding no evidence for a significant scaling between Galactic and SMC
metallicities for T$_{\rm eff} <~$30,000\,K, consistent with the
predictions of radiation driven wind theory.  A comparison of the
$v_\infty / v_{esc}$ ratio between the SMC and Galactic samples, while
consistent with the above statement, emphasizes that the uncertainties
in the distances to galactic OB-type stars are a serious obstacle to a
detailed comparison with theory.  For the SMC sample there is
considerable scatter in $v_\infty / v_{esc}$ at a given effective
temperature, perhaps indicative of uncertainties in stellar masses.
\end{abstract}

\keywords{Galaxies: individual: Magellanic Clouds -- stars: fundamental parameters --
stars: winds, outflows -- stars: early-type}

\section{Introduction}
\label{intro}

Observations of individual early-type stars in the Small Magellanic
Cloud (SMC) facilitate studies of stellar evolution and radiatively
driven winds in a metal-poor environment.  Stellar winds are thought
to be driven by momentum transfer from metal line absorption; thus,
one would $expect$ global wind properties to be dependent on
metallicity ($Z$).  Partial confirmation of this is offered from
morphological consideration of SMC ultraviolet spectra in which
P-Cygni type profiles are weaker (or even absent) when compared with
Galactic analogs (e.g., Walborn et al., 2000).  Previous UV surveys
using the {\it International Ultraviolet Explorer} telescope ($IUE$;
Garmany \& Conti, 1985) and the {\it Hubble Space Telescope} ($HST$;
Walborn et al.  1995) concluded that terminal velocities
($v_{\infty}$) for early-type stars are on average lower in the SMC
than in their Galactic counterparts.  This finding has subsequently
been echoed by Kudritzki \& Puls (2000) and Puls, Springmann, \&
Lennon (2000), albeit based on the same observational data.  In
addition, a theoretical prediction by Leitherer, Robert, \& Drissen
(1992) that $v_\infty$ scales as $Z^{0.13}$ is at times referred to in
the literature as having been confirmed (e.g., Vink, de Koter, \&
Lamers 2001).

A detailed analysis of the line driving process for radiatively driven winds, 
including its dependence on metallicity, was given by Puls et al. (2000).
They note that, to first order,
\begin{equation}
 v_\infty / v_{esc} \sim \frac{\hat{\alpha}}{1-\hat{\alpha}},
\end{equation}
where $\hat{\alpha}$ is the effective value of the force
multiplier parameter, $\alpha$, defined as the exponent of the line
strength distribution function.  This parameter depends on both
metallicity and depth (temperature and density) and Puls et al. give
explicit examples at various effective temperatures and metallicities.
Provided that we do not enter the regime of weak winds, their Figure
27 illustrates that for O-stars (T$_{\rm eff} = 40,000$\,K) the value
of $\hat{\alpha}$ is only weakly dependent on metallicity, at least
down to one tenth solar.  This is confirmed by the calculations of
Kudritzki (2002), who parameterized the temperature and density
variations of the force multiplier parameters to construct wind models
for O-type stars to very low metallicity (0.0001 $Z_{\odot}$).  From
Kudritzki's Table 2 and Figure 9 we see that, at the metallicity of
the SMC, one does not expect any significant change in the value of
$v_\infty$/$v_{esc}$ for very luminous O-type stars (in fact, for the
lowest surface gravity models there is even a small increase in
terminal velocity as metallicity is changed from solar to one-fifth
solar).  This result does not necessarily contradict Leitherer et al.,
who note that their scaling is a description of the average
behavior and that ``$v_\infty$ may increase with decreasing $Z$ in
certain $T_{\rm eff}$ domains'' as a result of varying ionization conditions in
the wind.  For the Kudritzki main-sequence models there is indeed a
weak trend with $Z$, which essentially agrees with the Leitherer et
al. scaling (which for one-fifth solar implies a change of only
20$\%$, comparable to the uncertainty arising from the determination
of the surface gravities).

In part prompted by the Kudritzki results, here we attempt a more
quantitative investigation of terminal velocities for supergiants in
the SMC; we determine $v_\infty$ for a total of 16 targets.  Aside
from consideration as a whole to study the effects of metallicity,
such results are useful individually.  Terminal velocities are a
necessary ``ingredient'' for the detailed atmospheric analyses now
possible for early-type stars.  Theoretical model atmosphere codes
have undergone significant development over the past decade, most
recently with the inclusion of line-blanketing arising from metallic
species e.g., {\sc cmfgen} (Hillier \& Miller, 1998) and {\sc fastwind}
(Santolaya-Rey, Puls, \& Herrero 1997; Herrero, Puls, \& Najarro 2002).  One of the most
salient results from the application of these codes to analyses of
O-type stars is the downward revision of stellar temperatures (when
compared with previous determinations that did not include the effects
of blanketing), noted for plane-parallel models by Hubeny, Heap, \& Lanz
(1998).  This effect has since been more thoroughly explored in
both the Galaxy and the Magellanic Clouds using extended atmospheres
e.g., Martins, Schaerer, \& Hillier (2002), Crowther
\ea\,(2002), Herrero \ea\,(2002) and Bouret et al. (2003).

Detailed line-blanketed methods have currently not been applied widely
to B-type stars.  In combination with high-resolution optical data
from the VLT UV-Visual Echelle Spectrograph (UVES), the $v_\infty$
results presented here for four of our early B-type Space Telescope
Imaging Spectrograph (STIS) targets are used by Trundle \ea~(2004,
hereafter TL04) to determine intrinsic stellar and wind parameters
using {\sc fastwind}.  Part of our motivation in addressing the
dependence of terminal velocities on environment also stems from the
TL04 study.  Ultraviolet spectra of four further SMC targets (observed
as part of the same STIS program) do not display evidence of strong
stellar winds and the methods we use here are not sufficient for
determination of $v_\infty$ (these additional data will be discussed
in a subsequent paper).  We therefore investigate the scaling relation
of $v_\infty$ with stellar escape velocity ($v_{\rm esc}$), with a
view to obtaining estimates of the terminal velocities for these
additional targets.  We note that such scaling relations are also of
interest as they are employed in the analysis of samples of stars
which are too distant or numerous to obtain UV spectra (e.g, Urbaneja
et al. 2003).

\section{Observations}
\label{obsdetails}
\subsection{$HST$-STIS observations}
In the course of our $HST$ General Observer programs (GO7437 and
GO9116, P.I.: DJL) we have observed a number of O and B-type SMC stars
that display evidence of strong stellar winds, for which terminal
velocities can be determined.  The dwarfs in the young cluster
NGC\,346 (observed as part of G07437) have recently been studied by
Bouret et al. (2003) and are not considered here.  Observational
details of our 11 targets are summarized in Table \ref{targets}; 
all are drawn from the catalog of Azzopardi \& Vigneau (AzV; 1975, 1982).

Each target was observed using the STIS in echelle mode, with the
far-UV MAMA (Multi-Anode Microchannel Array) detector and the E140M
grating (centered at 1425\,\AA); those from program GO9116 were also
observed with the near-UV MAMA and the E230M grating (centered at
1978\,\AA).  All of the observations were made through the
$0{\farcs}2~\times~0{\farcs}2$ entrance aperture and the effective
spectral resolving power of the E140M and E230M gratings are $R =
46,000$ and 30,000, respectively.  The exposure times for the GO9116
data were set from consideration of low-resolution $IUE$ spectra.
With the E140M grating the exposure times range from 54 to 216 minutes
(Sk\,191 and AzV\,18, respectively), with shorter exposures for the E230M set-up
(ranging from 38 to 75 minutes).

The primary {\sc calstis} processing steps are described by Walborn et
al. (2000), with the difference here that the two-dimensional
inter-order background correction performed manually in those
reductions has now been incorporated into the standard pipeline as
{\sc sc2dcorr}.  After pipeline reduction the individual echelle
orders were extracted and merged to form a continuous spectrum.  Prior
to merger (typically) 25 pixels at both ends of each order are
clipped; the spectral overlap between orders is generally large enough
to accommodate this and the final result has a more consistent
signal-to-noise ratio than otherwise.  The reduced echelle spectra cover the
wavelength ranges 1150-1700\,\AA\,(E140M) and 1600-2350\,\AA\,(E230M).
For display purposes (and for consistency with Walborn et al. 2000)
the spectra are binned to 0.25\,\AA~per data point.  The S/N (on the
basis of photon statistics) is typically in the range of 20-30 per
binned data point.

In Table \ref{targets} we also give H~\1 column densities, $N$(H~\1),
toward our targets, found assuming the Ly$\alpha$ profile arises
from a pure damping profile (see e.g., Shull \& Van Steenberg 1985 and
references therein).  To avoid possible contamination by the N~\5
\lam\lam1239, 1243 doublet, we consider only the blueward wing when
deriving the tabulated values.

\subsubsection{Multiplicity of AzV~47 and 216}

\begin{figure*}
\begin{center}
\begin{minipage}{4cm}
\includegraphics[width=4cm, height=3.5cm]{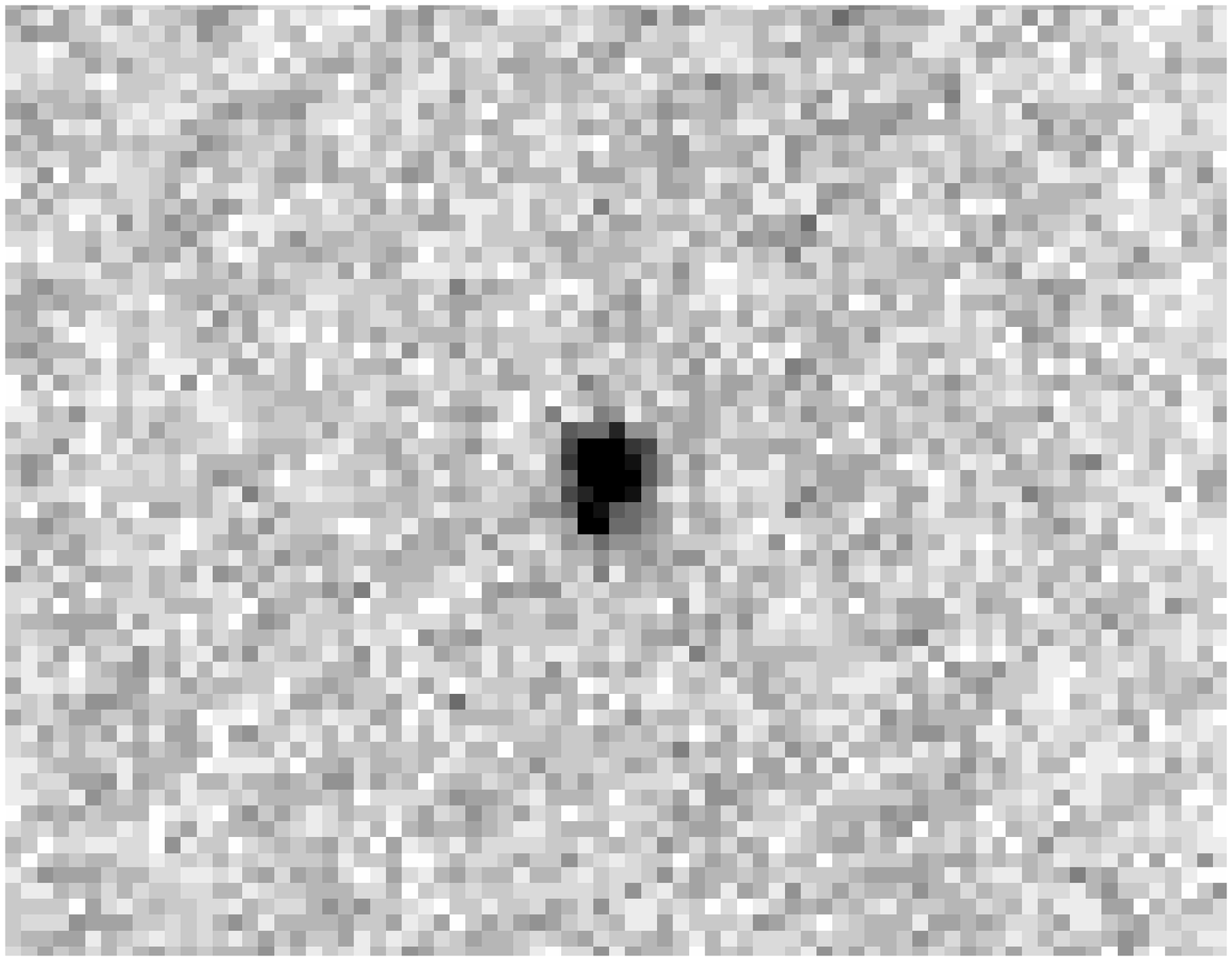}
\end{minipage} 
\hspace*{0.5cm}
\begin{minipage}{4cm}
\includegraphics[width=4cm, height=3.5cm]{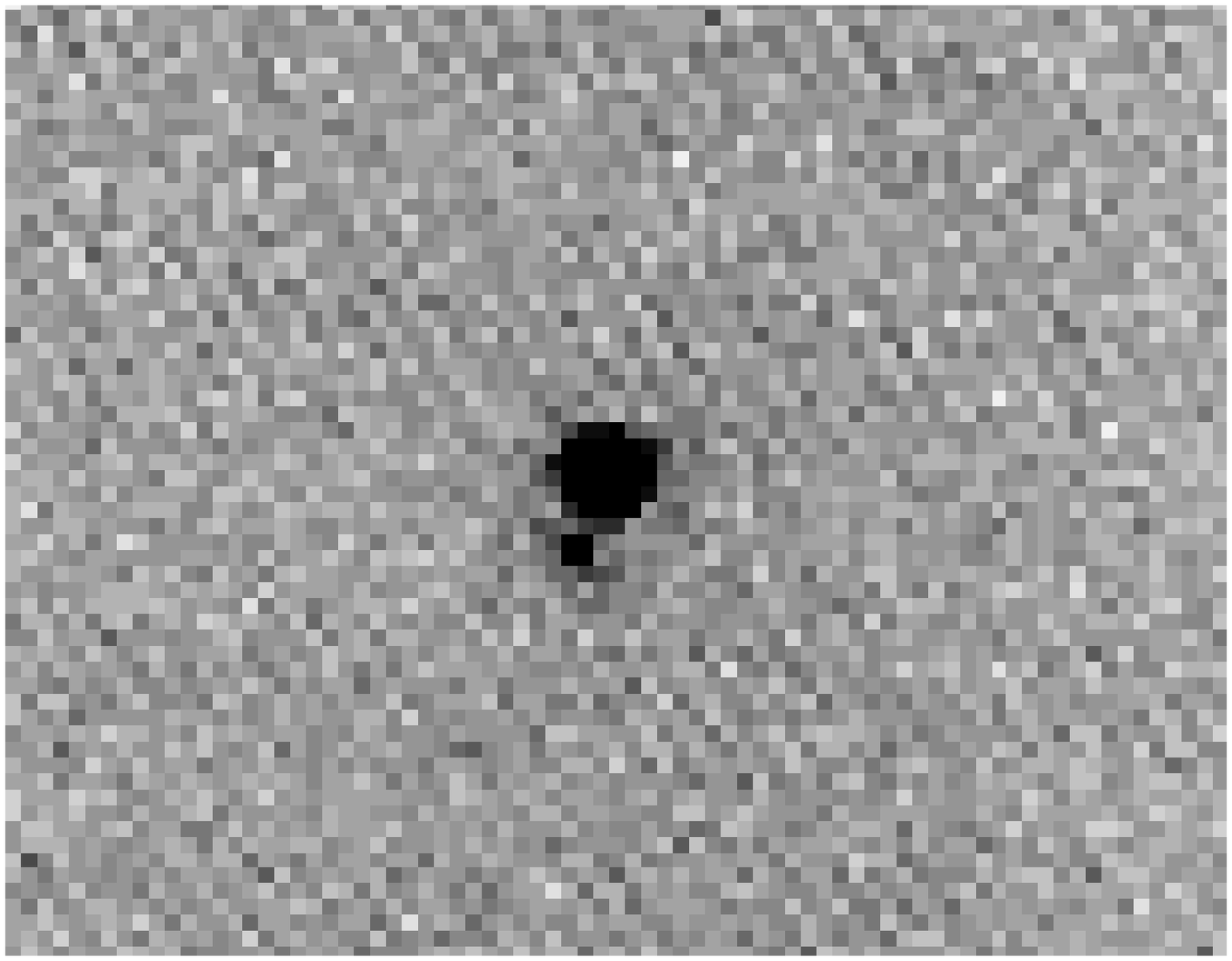}
\end{minipage}
\caption{STIS acquisition images for AzV~47 (left) and AzV~216 (right).  The
regions shown correspond to approximately 3{\farcs}8$\times$2{\farcs}9.\label{acq_images}}
\end{center}
\end{figure*}

Target acquisition images used for the STIS are taken using a
100$\times$100 pixel CCD camera (covering 5\arcsec$\times$5\arcsec),
with the ``longpass'' broad-band filter centered at
7230\,\AA~(FWHM~=~2720\,\AA).  Inspection of these images for each of
our targets generally show isolated point sources, with the exception
of a faint companion present in the image of AzV\,47.  While inspecting the
acquisition data from the additional targets observed for program
GO9116,\footnote{Lower luminosity targets were also observed for
compilation of spectral libraries and their UV spectra are not
discussed here.} it came to our attention that the acquisition image
for AzV\,216 (classified as B1\,III by TL04) also displays a faint
companion. These images are shown in Figure \ref{acq_images}.

The separations are $\sim$3 pixels (0.15$''$) in AzV 47 and $\sim$5
pixels (0.25$''$) in AzV 216 which, at the distance of the SMC,
corresponds to approximately 9000 and 15,000 AU, respectively; both
companions are relatively faint in comparison to the primary targets.
Some contamination is possible in the UV spectrum of AzV\,47 because
the companion will not be excluded by the
0{\farcs}2~$\times$~0{\farcs}2 aperture.  However, Walborn et
al. (2000) noted that ``the optical and UV spectra of AzV 47 are
entirely normal for its type,'' so it is likely that there is not a
significant contribution from the companion.  A full analysis of the
VLT-UVES spectrum of AzV\,216 is presented by TL04; both H$\alpha$ and
the He~\1 \lam4471 line show some asymmetry, likely originating from
the presence of a nearby companion.

\subsection{Additional UV data}

To augment the $HST$-STIS spectra, we include four B1-type stars
observed with the $HST$ using the Goddard High-Resolution Spectrograph
(GHRS) as part of program GO6078 (P.I.: DJL).  We also include a
high-resolution observation of AzV~78 made with the $IUE$ telescope 
using the large aperture with the short-wavelength prime (SWP) camera
(image SWP09334, Fitzpatrick \& Savage, 1985).  The GHRS observations
do not cover the Ly$\alpha$ region, and, although the spectrum of
AzV~78 extends far enough into the blue, the signal-to-noise in the
region of Ly$\alpha$ is not sufficiently high to permit
measurement of the H~\1 column density.  These additional targets are
also detailed in Table \ref{targets}.

\section{Determination of terminal velocities: SMC stars}
\label{sei}
The stellar winds of B-type stars are much less intense than those
observed in O-type stars; terminal velocities (e.g, Prinja, Barlow, \&
Howarth 1990) and
mass-loss rates (e.g., Kudritzki et al. 1999) are generally both lower in
B-type stars.  P Cygni--like profiles are seen in the Si~\4 and C~\4 resonance
lines of three stars (AzV\,210, 215, and Sk\,191), and
weak Si~\4 emission is seen in AzV\,18.  Terminal wind velocities
were determined for these four stars using the Sobolev with
Exact Integration (SEI) computer program developed by Haser (1995).
This program was then employed to find $v_\infty$-values for the seven O-type
STIS targets in Table \ref{targets} and for the five archival
GHRS/$IUE$ B-type spectra.  The same SEI program has been widely used
in other recent analyses of early-type stars in the Local Group,
e.g., Herrero et al. (2001, Milky Way), Bresolin \ea~(2002, M31), and
Urbaneja \ea~(2002, M33).

The SEI code adopts a standard $\beta$-law to describe the velocity
structure and includes a turbulent velocity, which increases with
radius to a maximum of $v_{\rm turb}$ at $v_{\infty}$ (see Haser et
al., 1995 for further details).  The C~\4 \lam\lam1548.19, 1550.76
and Si~\4 \lam\lam 1393.73, 1402.73 doublets are used as diagnostics
of $v_\infty$.  We do not use the N~\5 \lam\lam1238.81, 1242.80
doublet in our analysis as a consequence of its weakness in early
B-type spectra (which constitute the bulk of our sample) and because of
problems arising from its contamination by Ly$\alpha$ absorption.
The adopted $\beta$-law is essentially a free parameter in the SEI
program, to be determined from morphological comparisons of the
observed and theoretical resonance lines.  In the current
work we are solely interested in the terminal velocity and, though small
variations in $\beta$ can lead to more aesthetically pleasing
line fits, the resulting $v_\infty$-values are still within the quoted
errors.  Therefore, in contrast to other studies (e.g., Bresolin et al., 2002), we
do not attempt to ``fit'' the $\beta$-law for our targets and adopt
$\beta = 1$ (consistent with the results of e.g., Massa et al. 2003).

\begin{figure*}
\begin{center}
\includegraphics[scale=0.8]{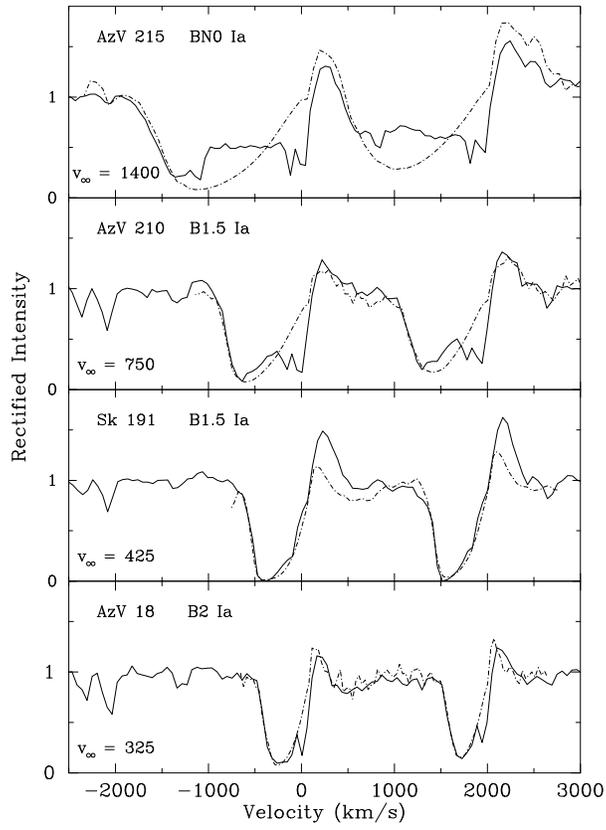}
\caption{Model fits to the Si~\4 doublet for STIS B-type spectra.
We are unable to match the morphology of AzV~215, though the same 
$v_\infty$ successfully matches the C~\4 profile (see Figure \ref{c_fits}).
\label{si_fits}}
\end{center}
\end{figure*}

Our results are given in Table \ref{targets}, and model fits for four
of our B-type STIS targets are shown in Figure \ref{si_fits}.  To
correct for the underlying photospheric components of the resonance
lines in our analysis, we use archival $IUE$ observations of B-type
stars with low projected rotational velocities (selected from Siebert,
1999) as templates,\footnote{Additional tests with new SMC metallicity
{\sc tlusty} models (R. Ryans, 2003, private communication)  yield essentially
identical terminal velocities.} convolved by the appropriate
\vsini~of each target (for estimates see TL04 and Walborn et al. 2000).
The model spectra were also convolved with an instrumental broadening 
function, although given the high resolution of these data this effectively
has no impact on the derived velocities.

\begin{figure*}
\begin{center}
\includegraphics[scale=0.8]{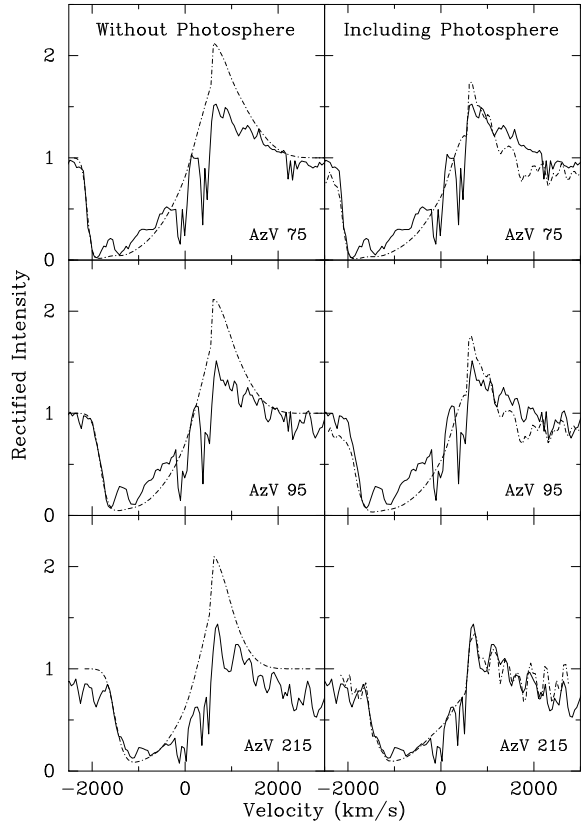}
\caption{Model fits to the C~\4 doublet for AzV 75 [O5 III(f$+$)], 
AzV 95 [O7 III((f))] and AzV 215 [BN0 Ia].  Models were calculated without
consideration of the photospheric contribution (left panel) and with 
inclusion of a photospheric template (right panel).  The derived terminal
velocities are identical.\label{c_fits}}
\end{center}
\end{figure*}

Terminal velocities for the B-type spectra are determined primarily
from consideration of the Si~\4 doublet.  However, as shown in the top
panel of Figure \ref{si_fits}, we are unable to match the morphology
of the Si~\4 doublet in AzV~215.  The presence of the relatively
flat ``shelf'' resembles that seen in $\kappa$~Ori (HD~38771, B0.5 Ia,
Walborn, Nicols-Bohlin, \& Panek 1985) though in AzV~215 it is more extreme.  The C~\4
profile for this star is shown in Figure \ref{c_fits} and is well
matched by a $v_\infty$ = 1400 \kms~model.

In determining $v_\infty$ for our O-type spectra we use the C~\4
profile as the primary diagnostic; the Si~\4 doublet is a relatively
weak feature in the O-type stars studied here (see Walborn \ea~2000).
Correcting for the photospheric component is complicated in O-type
stars by the absence of templates uncontaminated by stellar winds (the
same problem is discussed by Massa \ea~2003).  In Figure \ref{c_fits}
we show model fits for two of our O-type spectra (AzV 75 and 95) and
also for AzV~215.  In the case of AzV~215 the same $v_\infty$-value is found
regardless of whether the photospheric correction is included (using
$\upsilon$ Ori, HD 36512, Walborn \ea~1985), i.e., the large terminal
velocity of the wind leaves the blue-edge relatively free of
photospheric contamination.  To test if this also holds true for our
O-type stars we used the $IUE$ spectrum of the subluminous dwarf
BD$+$75 325 (classified as O5p [Gould, Herbig, \& Morgan 1957] and used by Herrero
\ea, 2001, in analysis of their hottest stars) as a template for AzV 75
and 95.  As demonstrated in Figure \ref{c_fits}, the blue edge of the
C~\4 profile is again largely independent of the inclusion of a
photospheric template, although a more visually pleasing fit is achieved
to the emission profile.  For the remaining O-type stars we simply match
the blue edge, without including a photospheric template.

It is important to note that the determination of the best fit in all
our spectra is subjective.  In the case of saturated lines different
combinations of $v_\infty$ and $v_{\rm turb}$ can be used to match the
blue wing of the profiles; in nonsaturated, weaker lines the adopted
line-strength of each doublet is an additional parameter to be
considered.  Typical uncertainties in $v_\infty$ are $\pm$\,100\,\kms,
except in the case of the four B-type stars with $v_\infty < 500$\kms,
where it is in the region of $\pm$\,50\,\kms.

\subsection{Comparison with previous results}

Terminal velocities for three of our targets were determined by Prinja
(1987) from comparison of low-dispersion $IUE$ spectra with
theoretical profiles.  His value for AzV 242 (1000 \kms) is within our
quoted uncertainty, though his result for AzV 215 (1675 \kms) is
slightly larger.  More worrying is that his value for Sk 191 (850
\kms) is twice our result.  This prompted us to retrieve the relevant
$IUE$ spectrum from the archive and it neither displays P-Cygni
emission nor saturated absorption profiles.  From the new
higher quality STIS data, measurement of the blueward limit of the
saturated core of both the C~\4 and Si~\4 doublets (generally referred
to as $v_{black}$ in the literature) gives a velocity of 450
\kms, proscribing the larger result.

A more recent study of AzV\,242 (Prinja \& Crowther, 1998) gives
$v_\infty = 860 \kms$ from consideration of observed narrow
absorption components  (NACs; a method we do not pursue here).  Albeit
at the lower limit, this result is again within our expected uncertainty.

Garmany \& Fitzpatrick (1988) also used low-dispersion $IUE$ spectra
to study a sample of SMC stars and there is some overlap with our
targets (namely AzV 15, 69, 75, and 95).  They were limited to estimates
of the terminal velocities, in effect measuring a $v_{edge}$ (usually
defined as where the absorption component of the P-Cygni profile meets
the continuum).  We do not consider such measurements here; the
relation of $v_{edge}$ to $v_\infty$ is dependent on the morphology of
the blueward part of the profile, which is primarily determined by the
turbulent velocity.  As one would expect, their velocities are
considerably larger ($\sim$500 \kms) than those from the new analyses.

Finally, detailed non-LTE, line-blanketed model atmosphere analyses of AzV\,69
and 83 were presented by Hillier et al. (2003).  From fitting the
absorption and emission line widths of selected UV lines, terminal
velocities of 1800\,\kms (AzV\,69) and 940\,\kms (AzV\,83) were obtained
from comparisons with model spectra; these results are well within the
quoted uncertainties when compared with our new SEI results in Table
\ref{targets}.

\section{Determination of terminal velocities: Galactic stars}
To provide a test of our methods we have also used the SEI program to
determine terminal velocities for five Galactic B-type supergiants
taken from the Walborn et al. (1985) $IUE$ atlas.  These stars are
drawn from the sample of $\sim$60 early B-type supergiants analysed by
Haser (1995), and a comparison of results is given in Table
\ref{results2}.  The velocities determined here are $smaller$ than
those found by Haser, and consequently the adopted turbulent velocities
are larger (to match the blue edge of each line as described in
\sref{sei}).

In their analyses of O-type stars Groenewegen \& Lamers (1989), Haser
(1995), and Massa \ea~(2003) adopted turbulent velocities in the range
0.05-0.15\,$v_\infty$.  Comparable results are found for the O-type
stars studied here, with the exception of AzV\,83.  Haser used similar
turbulent velocities for his B-type sample
(i.e., 0.05-0.15\,$v_\infty$).  This highlights a potential difference
in methods between Haser's and the current study; our adopted
turbulent velocities are generally larger.  Both Bresolin et
al. (2002) and Urbaneja et al. (2002) also noted that $v_{\rm
turb}$/$v_{\infty}$ was larger for their analyses of B-type spectra,
compared with previous O star results.

\begin{figure*}
\begin{center}
\includegraphics[scale=0.8]{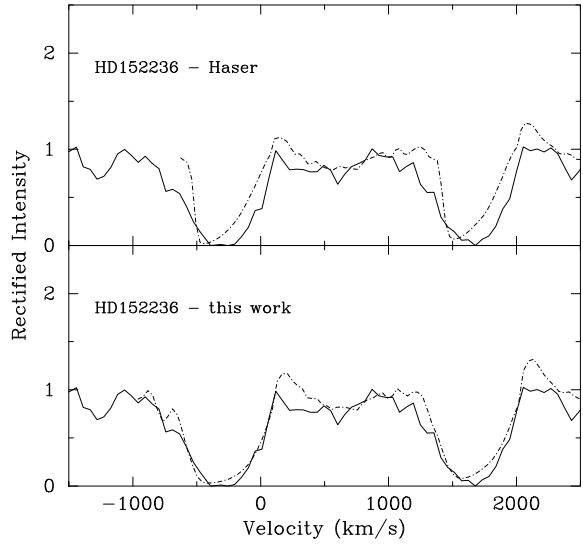}
\caption{Si~\4 doublet of HD152236 ({\it solid lines}) compared with model
profiles from the SEI code ({\it dot-dashed line}).  The top panel shows a model
calculated for $v_\infty = 500$ \kms, $v_{\rm turb} = 50$ \kms and $\beta = 1.5$ 
(Haser, 1995).  The lower panel shows the model from the current study:
$v_\infty = 450$ \kms, $v_{\rm turb} = 175$ \kms and $\beta = 1.0$.  All profiles
are shown in velocity space, relative to the blue component of the doublet, 
i.e., 1393.7 \AA.\label{haserfit}}
\end{center}
\end{figure*}

One factor contributing to the larger ratio is simply that the
terminal velocities in the cooler B-type stars are typically smaller
than those in the O stars.  Coupled with the fact that $v_{\rm turb}$
does not appear to vary systematically with spectral type this leads
to a larger $v_{\rm turb}$/$v_{\infty}$ ratio.  However, inspection of
Table \ref{results2} highlights an important issue that arises from
our Galactic analyses.  In all five cases we adopt larger turbulent
velocities than Haser to match the same observations, using the same
program.  The situation is most severe in HD\,152236, where $v_{\rm
turb}$ is a factor of 4 larger.  In Figure \ref{haserfit} we
compare our fit to the HD\,152236 Si~\4 doublet with a model calculated
using the parameters found by Haser.  The larger turbulent velocity
adopted in our study is clearly more consistent with the blue-edge of
each component than in the Haser model; note also that the width of
the absorption profile is better matched by our model.  The same
features are present when comparing the other four $IUE$ targets with
models calculated with Haser's values.  The Haser $v_\infty$ results
are generally at the ``upper limit'' of our results, suggesting that his
methods were slightly different; i.e., he found the maximum velocity
consistent with the data, whereas our approach has been to find the
range of $v_\infty$ (for which reasonable fits are found) and then
taking the central value.  These results highlight the dependence of
the derived terminal velocity on the adopted turbulence.  We emphasize
that $v_{\rm turb}$ in the present context is used very much as a
fitting parameter, the physical interpretation of such large
turbulent velocities is extremely uncertain.

As an aside, we note that the $IUE$ image used for our analysis of
HD~91969 is not that from the Walborn et al. atlas (SWP06510).  Haser
did not specify which image was used in his analysis, but his
$v_\infty$ of 1550\,\kms\,is comparable with previous results from
SWP09076, i.e., 1545 \kms\,(Prinja et al., 1990) and 1500 \kms\,(Lamers, 
Snow, \& Lindholm 1995).  The wind variability of HD~91969 has been discussed
recently by Prinja, Massa, \& Fullerton (2002) and is highlighted by our results;
for the SWP09076 spectrum we find $v_{\infty} = 1500$\,\kms, cf.~1350
\,\kms\,from SWP06510.

\section{Discussion}
\label{discuss}
The correlation between terminal and stellar escape velocities was
noted by Abbott (1978) and subsequent studies (e.g., Groenewegen, 
Lamers, \& Pauldrach, 1989; Howarth \& Prinja, 1989; Lamers \ea~1995) have revisited
this relationship.  A reliable, well-constrained $v_\infty$/$v_{esc}$
relation is an extremely attractive prospect.  For instance, if the
wind momentum-luminosity relationship (WLR, e.g., Kudritzki \& Puls
2000) is to be used to determine distances to galaxies beyond the
Local Group, ultraviolet stellar spectra are generally not available
(e.g, Urbaneja et al. 2003).  In comparison with theoretical spectra,
optical data can be used to provide an initial estimate of $v_{esc}$,
thence an estimate of $v_\infty$, with some iteration to a final
solution.  Before investigating this relationship for our SMC sample,
we compare our $v_\infty$ results with those from Galactic studies.

\subsection{Terminal velocities}
\label{vinf}
Common belief holds that terminal velocities in the SMC will be lower
than in the Galaxy because of the decrease in metallicity
(e.g., Kudritzki \& Puls, 2000).  Kudritzki \& Puls presented a
comparison of terminal velocities (as a function of temperature) for
Galactic and Magellanic Cloud targets.  A necessary step in this
process was the adoption of a suitable calibration to allocate
effective temperatures to a given spectral type (in their case that of
Humphreys \& McElroy, 1984).  Over the past 20 years, significant
developments have been made in model atmosphere techniques, most
recently the downward revision of the temperature scale discussed
earlier owing to the inclusion of line blanketing (e.g., Martins
\ea~2002).  If we are to compare our $v_\infty$-values with Galactic
results in this manner, it is clear that any differences in temperature
scales should be minimized.

\begin{figure*}
\begin{center}
\includegraphics[scale=0.8]{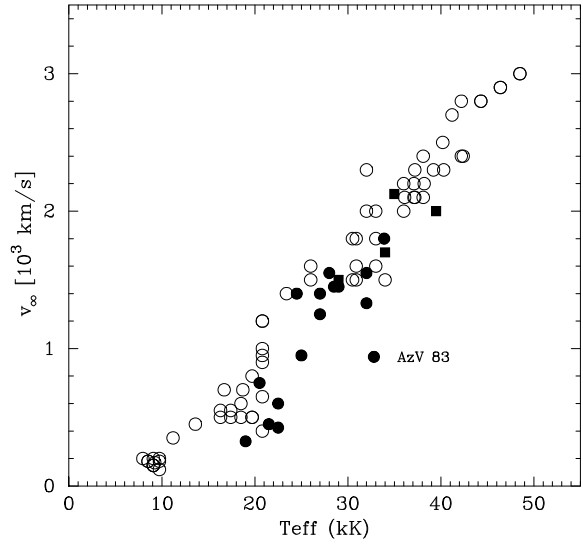}
\caption{Stellar terminal velocities, $v_\infty$, as a function of effective temperature
for the current SMC sample ({\it solid symbols}) and for the Galactic
targets of Lamers et al. (1995, {\it open symbols}).  Squares are those
stars for which the temperatures have been interpolated from published
results.  \label{vinf1}}
\end{center}
\end{figure*}

\begin{figure*}
\begin{center}
\includegraphics[scale=0.8]{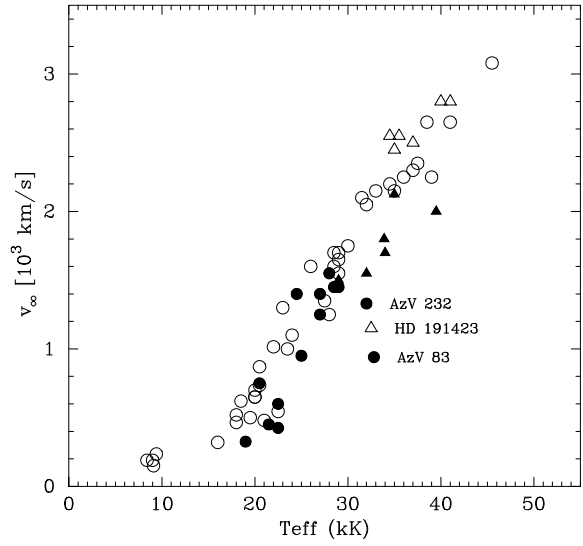}
\caption{Terminal velococity $v_\infty$ as a function of T$_{\rm eff}$ for the current SMC sample 
({\it solid symbols}) and for published analyses of Galactic targets ({\it open symbols}, 
see text for sources).  Supergiants are marked as circles, giants as triangles.\label{vinf2}}
\end{center}
\end{figure*}

Temperatures for the SMC targets are taken from line-blanketed,
non-LTE analyses by Hillier et al. (2003; AzV 69, 83), TL04 (Sk 191,
AzV 18, 210, 215), and Trundle et al. (2004, in preparation; AzV 78,
242, 264); note that for consistency with the analysis of AzV 69 and
83 by Hillier et al. we adopt their $v_\infty$-values.  Temperatures
for the remaining targets (AzV 15, 75, 95, 327)\footnote{We omit AzV
80 from this discussion since the exact nature of the peculiar ``nfp''
spectral class is unclear (Walborn et al. 2000).} are estimated from
interpolation between the line-blanketed results of Crowther et
al. (2002) and Martins et al.  (2002).  Our results are supplemented
by those from similar analyses of SMC targets by Crowther et
al. (2002; AzV 232) and Evans et al. (2004; AzV 70, 235, 372, 456,
469, 488).  Terminal velocities for this expanded sample are plotted
as a function of stellar effective temperature (T$_{\rm eff}$) in
Figures \ref{vinf1} and \ref{vinf2}.

In Figure \ref{vinf1} we compare our results with the data of Lamers
\ea~(1995).  Their O-type temperatures were taken from the unpublished
Garmany catalog and Chlebowski \& Garmany (1991), in which non-LTE
results were used to define a temperature scale.  Temperatures for
their B-type targets were taken from the LTE calibrations of
Schmidt-Kaler (1982).  From qualitative inspection of Figure \ref{vinf1}
it appears as though the SMC values are generally lower than the Galactic
values, as previously assumed in the literature.

In Figure \ref{vinf2} we show our results together with recent
results from detailed analysis of individual Galactic targets,
i.e., independent of any external temperature calibrations.
Results for O-type supergiants (and giants) are taken from the
non-LTE, line-blanketed studies of Herrero \ea~(2002) and Repolust, 
Puls, \& Herrero (2004).  TL04 presents such analyses
for B-type SMC supergiants, but there are no contemporary results for
Galactic B stars.  In Figure \ref{vinf2} we include results from
McErlean \ea~(1999), which employed non-LTE methods to analyze
Galactic B-type stars, but which predate the availability of
line-blanketed atmosphere codes; $v_\infty$ values are taken from
Haser (1995) or Howarth et al. (1997).  As discussed in TL04, for the
early B-type spectra the inclusion of line-blanketing has a relatively
minor effect on the temperature scale (cf.~the results of Dufton et al. 2000),
i.e., McErlean's temperatures are preferable to those assumed from
Schmidt-Kaler.

In contrast to the comparison with the Lamers \ea~sample, there is no
obvious $qualitative$ difference between the Galactic and SMC results
for T$_{\rm eff} < 30,000$\,K.  Linear fits to these results
(excluding the Galactic A-type supergiants) give slightly different
relations for the Galactic and SMC samples.  However, because of the
scatter of the results, these differences are not statistically
significant and therefore we do not undertake a rigorous quantitative
analysis.

At hotter temperatures the situation is somewhat more
ambiguous; our motivations for the current study were primarily
concerned with B-type supergiants and the sampling of targets with
T$_{\rm eff} > 30,000$\,K is relatively sparse.  In Figure \ref{vinf2}
we distinguish between luminosity class I and III objects.  With the
exception of the low velocity for HD\,191423, the Galactic giants
$appear$ to have generally larger $v_\infty$-values than the Galactic
supergiants.  A more objective comment is that $v_\infty$ for the SMC
giants is lower than for the Galactic targets (again excepting
HD\,191423), although from such a limited sample such differences
clearly cannot be attributed to the change in metallicity.  We are
similarly limited by our sample of more luminous O-type stars.  Both
AzV\,83 and 232 have low terminal velocities for their spectral types
(noted by Walborn et al. 2000) and appear distinctly different from
the Galactic supergiants.  However AzV\,83 and 232 are both ``Iaf'' type
spectra with dense winds (e.g., Hillier et al. 2003); $v_\infty$-values
for similarly extreme stars in the LMC are also relatively low, 
e.g., Sk~$-$66$^\circ$169 (Crowther et al. 2002, Massa et al. 2003).  The
Galactic supergiants in Figure \ref{vinf2} are less extreme objects
and therefore do not permit a sensible direct comparison of O-type
supergiant terminal velocities at the current time.

\subsection{Stellar escape velocities}
We now consider the escape velocities for SMC
stars with published intrinsic stellar parameters (see Table
\ref{vesc_res}).  As is usual in such studies, we define the {\it effective}
escape velocity ($v_{esc}$) as the gravitational escape velocity, corrected
for the contribution to the radiative acceleration caused by Thomson 
scattering by free electrons.  Therefore, 
\begin{equation}
\label{eqn1}
v_{esc} = \sqrt{\frac{2GM_{\ast}(1-\Gamma)}{R_{\ast}}}
\end{equation}
where $\Gamma$ is the ratio of the radiative acceleration from electron
scattering to the stellar gravity given by
\begin{equation}
\label{gamma}
\Gamma = \frac{\sigma_eL_{\ast}}{4\pi GM_{\ast}c}
\end{equation}
in which $\sigma_e$ is the electron scattering opacity, given by
$n_{e}\sigma_{\rm T}/\rho$, where $n_e$ is the electron number density,
$\sigma_{\rm T}$ is the Thomson cross-section, and $\rho$ is the
density of the gas.  From the TL04 {\sc fastwind} model for AzV~215,
$\sigma_e = 0.31$ (at $\tau$ = 2/3).  This value is consistent with
previous values in early-type stars (e.g., using equation \eref{gamma}
and their values for $\Gamma$, $\sigma_e =$ 0.28--0.32 in the Galactic
study of Lamers \ea, 1995) and is used here in our calculations.

Stellar radii, masses, and luminosities are taken from the relevant
papers; note, however, that the SMC distance modulus adopted by Hillier
\ea~was 19.1 (cf.~18.9 in the other studies) which will affect the
luminosity of the final atmospheric models, hence the radii and masses
and ultimately the escape velocity (of order 10$\%$).  

\begin{figure*}
\begin{center}
\includegraphics[scale=0.8]{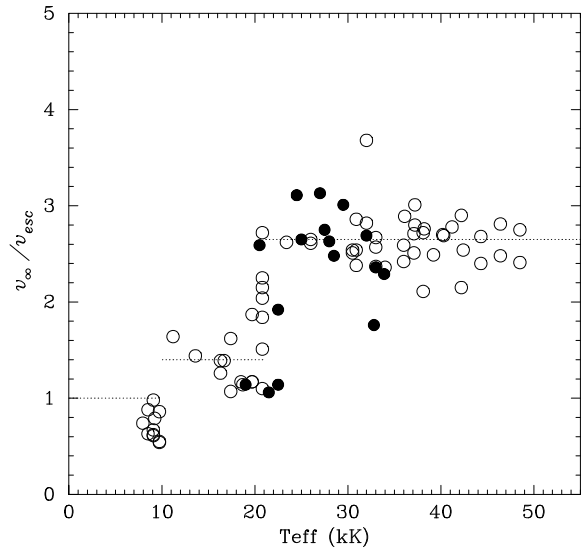}
\caption{Ratio $v_\infty/v_{esc}$ as a function of T$_{\rm eff}$ from the 
current SMC study ({\it solid circles}) compared with those from the
Galactic sample of Lamers \ea~(1995, {\it open circles}).  The scalings of 
Kudritzki \& Puls (2000) are also shown ({\it dotted lines}). \label{vesc1}}
\end{center}
\end{figure*}

In Figure \ref{vesc1} we compare $v_\infty$/$v_{\rm esc}$ for the SMC
stars in Table \ref{vesc_res} with the results from Lamers \ea~(1995).
As discussed before in \sref{vinf}, the temperature scales are not
directly comparable but both the scatter and the mean value of
$v_\infty$/$v_{\rm esc}$ is roughly consistent with that of the Lamers
sample.  From a compilation of published results, Kudritzki
\& Puls (2000) found $v_\infty$ = 2.65\,$v_{esc}$ (T$_{\rm eff} \ge$
21,000 K) quoting an accuracy of $\sim$\,20\,$\%$.  The mean from
our sample (excluding AzV~18 and 210 as they are cooler than 21,000 K)
is 2.36; AzV\,83 is again the most distant outlier.  

\begin{figure*}
\begin{center}
\includegraphics[scale=0.8]{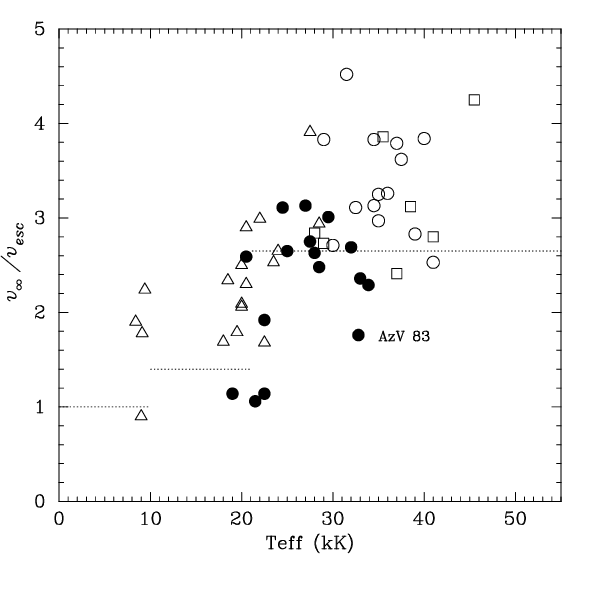}
\caption{Ratio $v_\infty/v_{esc}$ as a function of T$_{\rm eff}$ from the 
current SMC study ({\it solid circles}) compared with those from the
Galactic samples of Kudritzki et al. (1999, {\it open triangles}), Herrero
et al. (2002, {\it open squares}) and Repolust et al.  (2004, open circles).
The scalings of Kudritzki \& Puls (2000) are again shown 
({\it dotted lines}).  \label{vesc2}} 
\end{center} 
\end{figure*}

In Figure \ref{vesc2} our results are compared with those for Galactic
O-type giants/supergiants (Herrero et al. 2002; Repolust et al. 2004)
and B and A-type supergiants (Kudritzki et al. 1999).  The results for
HD\,207198 and 209975 (Repolust et al. 2004; $v_\infty$/$v_{\rm esc} =
6.09$ and 5.62, respectively) are excluded from the figure.  These two
stars are listed by Markova et al. (2004) as being members of the Cep
OB2 association, for which they adopt a distance of 0.85 kpc.
However, {\it Hipparcos} results by de Zeeuw et al. (1999) support
a smaller distance of 0.62 kpc. The region is also extended over
several degrees, so there will likely be a significant depth effect
within the association, leading to uncertain relative distances.
From equation (\ref{eqn1}) overestimates of their luminosities will
lead to underestimates of $v_{esc}$.  Smaller distances for these two
stars would bring them into better agreement with the other results in
Figure \ref{vesc2}.  The mean $v_\infty$/$v_{esc}$ of the remaining Galactic
targets (T$_{\rm eff} \ge$ 21,000 K) is 3.07, more in keeping with
Abbot's (1978) suggestion that $v_\infty \sim 3v_{esc}$ than the
Kudritzki \& Puls result.

Uncertainties in the distances to the Galactic targets may contribute
to these differing results and also to the greater spread of
$v_\infty$/$v_{\rm esc}$ (cf.\,Lamers et al.).  However, it is worth
noting that the targets in the Herrero et al. (2002) study are all
(thought to be) members of Cyg OB2;  aside from any systematic
offset, the spread of results for their seven targets should be
minimized and yet a wide range of $v_\infty$/$v_{\rm esc}$ is still
found.  Unfortunately, from current methods there will always be a
large scatter associated with $v_\infty$/$v_{esc}$, primarily a result
of the uncertainty in $\log g$ of (at least) $\pm$\,0.1 dex.  This
corresponds to a large uncertainty in the spectroscopic mass, which
in the case of, e.g., AzV~215, yields results for $v_{esc}$ of $\pm$\,20\,$\%$.

In addition to its use for targets without ultraviolet information,
understanding the scaling of $v_\infty$ with $v_{esc}$ is useful for
targets such as AzV\,22 and 362 (see TL04 for analysis).  The optical
spectra of these two stars display evidence of a stellar wind, but
their ultraviolet spectra (also observed as part of GO9116) are devoid
of strong wind signatures. Similarly, infilling of the H$\alpha$
profiles is observed in AzV\,104 and 216 indicative of some, albeit
weak, stellar wind.  Since we are limited in the current work by a
relatively small sample, TL04 used the Kudritzki \& Puls formula to
obtain $v_\infty$ estimates for inclusion in {\sc fastwind}.

In general, the results in Figure \ref{vesc2} display no evidence for
a significantly different $v_\infty$/$v_{\rm esc}$ ratio in the SMC in
comparison with that for Galactic stars.  We therefore conclude that in the case
of supergiants with strong winds, our results are in good agreement
with predictions of theory.

Figures \ref{vesc1} and \ref{vesc2} also prompt further discussion in
the context of the so-called ``bistability jump'' near T$_{\rm eff}
=$\,21,000\,K (Lamers et al. 1995).  From Figure \ref{vesc1}, there
appears to be a clear step between the high (hotter) and low (cooler)
ionization regimes, but such a well-defined edge is largely an artifact
of the temperature assignment for the spectral bins.  The Lamers et
al. sample includes seven B1-type stars for which the same temperature
(20,800\,K) is used; in reality, the temperature for a given spectral
type encompasses a range of values, e.g., Sk\,191 and AzV\,210 are both
B1.5 Ia-type spectra, yet their temperatures differ by 2000\,K
(TL04).  Further complications can arise from the effects of line-blanketing, 
i.e., the temperature
derived for a star with a strong wind can be lower than that found for
a later spectral type with a much weaker wind.  In these cases the
temperature scale is not necessarily monotonic (e.g., TL04, Evans et
al. 2004) and the temperature range for a given spectral type can
overlap with that of another.  In Figure
\ref{vesc2} there is still evidence for a lower $v_\infty$/$v_{\rm
esc}$ ratio at cooler temperatures, but, as a consequence of analysing
individual stars rather than employing an effective
temperature-spectral type calibration, the boundary is somewhat
``blurred.''  Additionally, we note that the median $v_\infty$/$v_{\rm
esc}$ for those SMC stars significantly hotter than this region
(T$_{\rm eff} >$ 24,000\,K) is 2.63, consistent with the Kudritzki \&
Puls relation.

\section{Conclusions}
From high-quality ultraviolet spectra we have used the SEI method to
determine terminal velocities for nine B-type supergiants and seven
O-type stars in the SMC.  One important conclusion from the current
work is that, if one is solely interested in terminal velocities, many
of the parameters in the SEI code are essentially cosmetic fitting
parameters to which other authors have attributed undue significance
in previous studies.

We compare our results with those from recent Galactic studies,
adopting temperatures from tailored atmospheric analyses, independent
of external calibrations.  To our knowledge, this is the first time
such a thorough comparison has been made, giving a clearer view of the
scaling of terminal velocities.  We find no evidence for lower
terminal velocities in the SMC than in the Galaxy for T$_{\rm eff}
<$\,30,000\,K (corresponding to spectral types later than O9).
Similarly, in agreement with the predictions of theory for luminous
supergiants, we find no evidence for a significantly different scaling
of $v_{\infty}$ with $v_{\rm esc}$ in the SMC in comparison with that in
the Galaxy.  We stress that the situation for giants and dwarfs remains
untested.

From our limited sample it appears that $v_\infty$ for O-type
giants may be lower in the SMC than in the Galaxy, although both this
and the situation for O-type supergiants warrants further
investigation.  Further results from contemporary model atmosphere
analyses of O-type stars in the Galaxy and the Magellanic Clouds would
greatly help attempts to delineate the scaling of $v_\infty$ with both
temperature and $v_{esc}$.  

Finally, given the uncertain distances, it would appear that
calibrating the WLR using Galactic OB-type stars is not an attractive
prospect at the current time.

\section{Acknowledgements}
CJE (under grant PPA/G/S/2001/00131) and DJL acknowledge financial
support from the UK Particle Physics and Astronomy Research Council
(PPARC).  CT is grateful to the Department of Higher and Further
Education, Training, and Employment for Northern Ireland (DEFHTE) and
the Dunville Scholarships fund for their financial support.  Based in
part on observations with the NASA/ESA {\it Hubble Space Telescope}
obtained at the Space Telescope Science Institute, which is operated
by the Association of Universities for Research in Astronomy, Inc. and
on INES data from the $IUE$ satellite.  We thank Miguel Urbaneja for
his useful initial assistance with the SEI program and Robert Ryans
for the {\sc tlusty} models.  We also thank both Alex Fullerton and the
referee for their helpful comments on the manuscript.

\clearpage

\begin{deluxetable}{lllccccllcccc}
\rotate
\tabletypesize{\scriptsize}
\tablewidth{0pc}
\tablecolumns{13}
\tablecaption{Observational parameters of SMC targets \label{targets}}  
\tablehead{
\colhead{Star} & \colhead{Alias} & \colhead{Sp. Type} & \colhead{$V$} & \colhead{$B - V$} & \colhead{Ref.} 
& \colhead{$v_r$} & \colhead{Instrument} & \colhead{Program} & \colhead{log $N$(H\,{\tiny I})} & \colhead{$v_\infty$}
& \colhead{$v_{\rm turb}$} & \colhead{$v_{\rm turb}$/$v_\infty$} \\ 
& & & & & & \colhead{[\kms]} & & & \colhead{[cm$^{-2}$]} & \colhead{[\kms]} & \colhead{[\kms]} & }
\startdata
AzV 80     & --     & O4-6n(f)p       & 13.33 & $-$0.14      & 1 &  131 & STIS        & GO7437   &  21.7  &      1550 &  250 & 0.16 \\
AzV 75     & Sk 38  & O5 III(f$+$)    & 12.79 & $-$0.16      & 1 &  141 & STIS        & GO7437   &  21.7  &      2000 &  175 & 0.09 \\
AzV 15     & Sk 10  & O6.5 II(f)      & 13.20 & $-$0.22      & 1 &  115 & STIS        & GO7437   &  21.5  &      2125 &  175 & 0.08 \\
AzV 83     & --     & O7 Iaf$+$       & 13.58 & $-$0.13      & 4 &  118 & STIS        & GO7437   &  21.1  & \ph{1}925 &  225 & 0.24 \\
AzV 95     & --     & O7 III((f))     & 13.91 & $-$0.30      & 1 &  137 & STIS        & GO7437   &  21.5  &      1700 &  250 & 0.15 \\
AzV 69     & Sk 34  & OC7.5 III((f))  & 13.35 & $-$0.22      & 1 &  130 & STIS        & GO7437   &  21.6  &      1750 &  150 & 0.09 \\
AzV 327    & --     & O9.5 II-Ibw     & 13.25 & $-$0.22      & 1 &  177 & STIS        & GO7437   &  21.0  &      1500 &  200 & 0.13 \\
AzV 215    & Sk 76  & BN0 Ia          & 12.69 & $-$0.09      & 3 &  157 & STIS        & GO9116   &  21.8  &      1400 &  275 & 0.20 \\
AzV 242    & Sk 85  & B1 Ia           & 12.11 & $-$0.13      & 1 &  176 & GHRS        & GO6078   &   --   & \ph{1}950 &  175 & 0.18 \\
AzV 264    & Sk 94  & B1 Ia           & 12.36 & $-$0.15      & 1 &  131 & GHRS        & GO6078   &   --   & \ph{1}600 &  250 & 0.42 \\   
AzV 340    & --     & B1 Ia           & 12.69 & $-$0.07      & 3 &  211 & GHRS        & GO6078   &   --   & \ph{1}975 &  150 & 0.15 \\   
AzV 78     & Sk 40  & B1 Ia$+$        & 11.05 & $-$0.03      & 2 &  164 & $IUE$-HIRES & SWP09334 &   --   & \ph{1}450 &  200 & 0.44 \\  
AzV 483    & Sk 156 & B1.5 Ia$+$      & 11.85 & $-$0.08      & 1 &  185 & GHRS        & GO6078   &   --   & \ph{1}400 &  250 & 0.63 \\   
Sk 191    & --     & B1.5 Ia         & 11.86 & $-$0.04      & 2 &  130 & STIS        & GO9116   &  21.4  & \ph{1}425 &  125 & 0.29 \\
AzV 210    & Sk 73  & B1.5 Ia         & 12.60 & $-$0.02      & 3 &  173 & STIS        & GO9116   &  21.6  & \ph{1}750 &  150 & 0.20 \\
AzV 18     & Sk 13  & B2 Ia           & 12.46 & \ph{-}0.03   & 2 &  138 & STIS        & GO9116   &  21.8  & \ph{1}325 &  125 & 0.38 \\
\enddata
\tablerefs{(1) Azzopardi \& Vigneau 1975, 1982; (2) Garmany et al. 1987; (3) Massey 2002; 
(4) Walborn et al. 2000}
\tablecomments{Stellar identifications are those of Sanduleak (1968, Sk) 
and Azzopardi \& Vigneau (1975, 1982, AzV).  Classifications for the O-type
spectra are from Walborn et al. (2000); B-type classifications are
from Lennon (1997), excepting AzV~483 (Lennon, 1999) and AzV~78 (from
re-inspection of the blue optical data the spectrum is now classified
as B1 Ia$+$).  Radial velocities are found from the blue optical data from
Lennon (1997) and an unpublished echelle spectrum of AzV~483 (using
CASPEC at the ESO 3.6-m telescope).  For AzV 18, 78, 483 and Sk~191
(i.e., $v_\infty <$ 500) the uncertainty in $v_\infty$ is $\pm$50 \kms
and for the remainder of the sample this increases to $\pm$100\,\kms.}
\end{deluxetable}

\clearpage

\begin{center}
\begin{deluxetable}{lllccccc}
\tablecolumns{8}
\tablewidth{0pc}
\tabletypesize{\small}
\tablecaption{Terminal velocities determined for selected Galactic B-type supergiants \label{results2}}
\tablehead{
 & & & \multicolumn{3}{c}{This work} & \multicolumn{2}{c}{Haser} \\
\colhead{Star} & \colhead{Spectral Type} & \colhead{$IUE$ image} &
\colhead{$v_{\infty}$} & \colhead{$v_{\rm turb}$} & \colhead{$v_{\rm turb}/v_\infty$} & \colhead{$v_{\infty}$} & 
\colhead{$v_{\rm turb}/v_\infty$}
}
\startdata
HD 91969  & B0 Ia      & SWP09076 &       1500 & 150 & 0.10 &      1550 & 0.09 \\
HD 115842 & B0.5 Ia    & SWP27405 &       1125 & 225 & 0.20 &      1200 & 0.13 \\ 
HD 148688 & B1 Ia      & SWP01871 &  \ph{1}625 & 175 & 0.28 & \ph{1}700 & 0.13 \\ 
HD 152236 & B1.5 Ia$+$ & SWP06500 &  \ph{1}450 & 175 & 0.39 & \ph{1}500 & 0.10 \\ 
HD 14818  & B2 Ia      & SWP09416 &  \ph{1}600 & 100 & 0.17 & \ph{1}650 & 0.10 \\ 
\enddata
\tablecomments{All tabulated velocities are given in \kms, typical uncertainties of
$v_{\infty}$ are $\pm$100 \kms.}
\end{deluxetable}
\end{center}

\clearpage

\begin{center}
\begin{deluxetable}{lccccc}
\tablecolumns{5}
\tablewidth{0pc}
\tablecaption{Escape velocities for O and early B-type SMC stars \label{vesc_res}}
\tablehead{
\colhead{Star} & \colhead{T$_{\rm eff}$} & \colhead{Ref.} & \colhead{$v_\infty$} & 
\colhead{$v_{esc}$} & \colhead{$v_\infty/v_{esc}$} \\
& \colhead{[kK]} & & \colhead{[\kms]} & \colhead{[\kms]} 
}
\startdata
AzV 69  & 33.9 & 3 &      1800 & \p785: & \p2.29: \\
AzV 469 & 33.0 & 2 &      1550 &    657 & 2.36 \\
AzV 83  & 32.8 & 3 & \ph{1}940 & \p534: & \p1.76: \\
AzV 232 & 32.0 & 1 &      1330 &    495 & 2.69 \\ 
AzV 456 & 29.5 & 2 &      1450 &    482 & 3.01 \\
AzV 70  & 28.5 & 2 &      1450 &    585 & 2.48 \\
AzV 372 & 28.0 & 2 &      1550 &    589 & 2.63 \\
AzV 215 & 27.0 & 4 &      1400 &    447 & 3.13 \\
AzV 488 & 27.5 & 2 &      1250 &    455 & 2.75 \\
AzV 242 & 25.0 & 5 & \ph{1}950 &    359 & 2.65 \\
AzV 235 & 24.5 & 2 &      1400 &    450 & 3.11 \\
AzV 264 & 22.5 & 5 & \ph{1}600 &    313 & 1.92 \\
Sk 191  & 22.5 & 4 & \ph{1}425 &    374 & 1.14 \\
AzV 78  & 21.5 & 5 & \ph{1}450 &    423 & 1.06 \\
AzV 210 & 20.5 & 4 & \ph{1}750 &    290 & 2.59 \\
AzV 18  & 19.0 & 4 & \ph{1}325 &    284 & 1.14 \\

\enddata
\tablecomments{Stellar temperatures have been determined using non-LTE,
line-blanketed model atmospheres.  The values for AzV~83 and 69 are
flagged as uncertain as a result of the larger SMC distance adopted in
their analysis compared with the other studies.}
\tablerefs{(1) Crowther \ea~2002; (2) Evans \ea~2004; (3) Hillier \ea~2003;
(4) Trundle \ea~2004, (5) Trundle \ea~2004, in preparation.}
\end{deluxetable}
\end{center}

\clearpage 

\clearpage

\clearpage

\clearpage

\clearpage

\clearpage

\clearpage

\clearpage

\begin{thebibliography}{100}
\bibitem{a78}
Abbott D. C., 1978, ApJ, 225, 893

\bibitem{av}
Azzopardi M. \& Vigneau J., 1975, A\&AS, 22, 285

\bibitem{av82}
Azzopardi M. \& Vigneau J., 1982, A\&AS, 50, 291


\bibitem{b03}
Bouret J.-C., Lanz T., Hillier D. J. et al., 2003, ApJ, 595, 1182

\bibitem{b02}
Bresolin F., Kudritzki R.-P., Lennon D. J. et al., 2002, ApJ, 580, 213

\bibitem{cg91}
Chlebowski, T., Garmany C. D., 1991, ApJ, 368, 241

\bibitem{c02}
Crowther P. A., Hillier D. J., Evans C. J. et al., 2002, ApJ, 579, 774


\bibitem{d99}
de Zeeuw P. T., Hoogerwerf R., de Bruijne J. H. J. et al., 1999, AJ, 117, 354

\bibitem{d00}
Dufton P. L., McErlean N. D., Lennon D. J., Ryans R. S. I., 2000, A\&A, 353, 311

\bibitem{e04}
Evans C. J., Crowther P. A., Fullerton A. W., Hillier D. J., 2004, ApJ, submitted



\bibitem{fs85}
Fitzpatrick E. L., Savage B. D., 1985, ApJ, 292, 122

\bibitem{g85}
Garmany C. D., Conti P. S., 1985, ApJ, 293, 407

\bibitem{g87}
Garmany C. D., Conti P. S., Massey P., 1987, AJ, 93, 1070

\bibitem{g88}
Garmany C. D., Fitzpatrick E. L., 1988, ApJ, 332, 711

\bibitem{g57}
Gould N. L., Herbig G. H., Morgan W. W., 1957, PASP, 69, 242

\bibitem{g89}
Groenewegen M. A. T., Lamers H. J. G. L. M., 1989, A\&AS, 79, 359

\bibitem{g89b}
Groenewegen M. A. T., Lamers H. J. G. L. M., Pauldrach A. W. A., 1989b, A\&A, 221, 78


\bibitem{h95}
Haser S. M., 1995, Ph.D. thesis, Univ. of Munich

\bibitem{hlk95}
Haser S. M., Lennon D. J., Kudritzki R.-P. et al., 1995, A\&A, 295, 136

\bibitem{h01}
Herrero A., Puls J., Corral L. J. et al., 2001, A\&A, 366, 623

\bibitem{h02}
Herrero A., Puls J., Najarro F., 2002, A\&A, 396, 949

\bibitem{hm98}
Hillier D. J., Miller D. L., 1998, ApJ, 496, 407

\bibitem{h03}
Hillier D. J., Lanz T., Heap S. R. et al., 2003, ApJ, 588. 1039

\bibitem{hp89}
Howarth I. D., Prinja R. K., 1989, ApJS, 69, 527

\bibitem{h97}
Howarth I. D., Siebert K. W., Hussain G. A. J., Prinja R. K., 1997, MNRAS, 284, 265


\bibitem{hhl}
Hubeny I., Heap S. R., Lanz T., 1998, in ASP Conf. Ser 131, Boulder-Munich II:
Properties of Hot, Luminous Stars, ed. I. D. Howarth (San Francisco, ASP), 108

\bibitem{hm84}
Humphreys R. M., McElroy, D. B., 1984, ApJ, 284, 565

\bibitem{k99}
Kudritzki R.-P., Puls J., Lennon D. J. et al., 1999, A\&A, 350, 970

\bibitem{k00}
Kudritzki R.-P., Puls J., 2000, ARA\&A, 38, 613

\bibitem{k02}
Kudritzki, R.-P., 2002, ApJ, 577, 389

\bibitem{l95}
Lamers H. J. G. L. M., Snow T. P., Lindholm D. M., 1995, ApJ, 455, 269

\bibitem{l92}
Leitherer C., Robert, C., Drissen, L., 1992, ApJ, 401, 596



\bibitem{l97}
Lennon D. J., 1997, A\&A, 317, 871

\bibitem{l99}
Lennon D. J., 1999, RMxAC, 8, 21


\bibitem{m04}
Markova N., Puls J., Repolust T., Markov H., 2004, A\&A, 413, 693

\bibitem{martins}
Martins F., Schaerer D., Hillier D. J., 2002, A\&A, 382, 999

\bibitem{massa}
Massa D., Fullerton A. W., Sonneborn G., Hutchings J.B., 2003, ApJ, 586, 996

\bibitem{m02}
Massey P., 2002, ApJS, 141, 81

\bibitem{mcer99}
McErlean N.D., Lennon D.J., Dufton P. L., 1999, A\&A 349, 553

\bibitem{p87}
Prinja R. K., 1987, MNRAS, 228, 173

\bibitem{p90}
Prinja R. K., Barlow M. J., Howarth I. D., 1990, ApJ, 361, 607

\bibitem{p98}
Prinja R. K., Crowther P.A., 1998, MNRAS, 300, 828

\bibitem{p02}
Prinja R. K., Massa D., Fullerton A. W., 2002, A\&A, 388, 587

\bibitem{p00}
Puls, J., Springmann, U., Lennon, M., 2000, A\&AS, 141, 23

\bibitem{r03}
Repolust T., Puls J., Herrero A., 2004, A\&A, 415, 349


\bibitem{sk}
Sanduleak N., 1968, AJ, 73, 246

\bibitem{fast}
Santolaya-Rey A. E., Puls J., Herrero A., 1997, A\&A, 323, 488


\bibitem{sk82}
Schmidt-Kaler T., 1982, in Schaifers K., Voigt H. H., eds., 
Landolt-B$\ddot{\rm o}$rnstein, Group VI, Vol 2b, Springer-Verlag, p.~1

\bibitem{sv85}
Shull J. M., Van Steenberg M. E., 1985, ApJ, 294, 599

\bibitem{kaj}
Siebert K. W., 1999, Ph.D. thesis, Univ. of London 

\bibitem{trundle03}
Trundle C., Lennon D. J., Puls J., Dufton P. L., 2004, A\&A, in press (TL04)



\bibitem{u02}
Urbaneja M. A., Herrero A., Kudritzki R.-P. et al., 2002, A\&A, 386, 1019

\bibitem{u03}
Urbaneja M. A., Herrero A., Bresolin F. et al., 2003, ApJ, 584, L73




\bibitem{v01}
Vink, J. S., de Koter, A., Lamers, H. J. G. L. M., 2001, A\&A, 369, 574

\bibitem{wnp85}
Walborn N. R., Nicols-Bohlin J., Panek R. J., 1985, 
International Ultraviolet Explorer Atlas of O-type Spectra from 1200 to 1900\AA, 
(NASA RP1363; Washington; NASA)

\bibitem{w95}
Walborn N. R., Lennon D. J., Haser S. M. et al., 1995, PASP, 107, 104

\bibitem{w00}
Walborn N. R., Lennon D. J., Heap S. R. et al., 2000, PASP, 112, 1243


\end{thebibliography}
\end{document}